\documentclass[prd,showpacs,amsmath,amssymb,twocolumn,superscriptaddress]{revtex4}
\usepackage{bm,multirow}
\usepackage{lipsum}
\usepackage{color}
\newcommand \beq{\begin{eqnarray}}
\newcommand \eeq{\end{eqnarray}}
\def\simge{\mathrel{%
       \rlap{\raise 0.511ex \hbox{$>$}}{\lower 0.511ex \hbox{$\sim$}}}}
\def\simle{\mathrel{
       \rlap{\raise 0.511ex \hbox{$<$}}{\lower 0.511ex \hbox{$\sim$}}}}
\usepackage{graphics}
\usepackage[dvips]{graphicx}
\usepackage{graphicx}

\newcommand{\be}{\begin{equation}}
\newcommand{\ee}{\end{equation}}
\newcommand{\bea}{\begin{eqnarray}}
\newcommand{\eea}{\end{eqnarray}}
\newcommand{\ba}{\begin{align}}
\newcommand{\ea}{\end{align}}

\newcommand{\rmi}{{\rm i}}

\begin{document}
\title{Snake instability of dark solitons in fermionic superfluids}
\author{A. Cetoli}
\author{J. Brand}     
\affiliation{New Zealand Institute for Advanced Study and Centre for
          Theoretical Chemistry and Physics, Massey University, Private Bag
          102904 NSMC, Auckland 0745, New Zealand}

\author{R.G. Scott}
\author{F. Dalfovo}
\affiliation{INO-CNR BEC Center and Dipartimento di Fisica, Universit\`a di Trento, Via Sommarive 14, I-38123 Povo, Italy}
\author{L.P. Pitaevskii}
\affiliation{INO-CNR BEC Center and Dipartimento di Fisica, Universit\`a di Trento, Via Sommarive 14, I-38123 Povo, Italy}
\affiliation{Kapitza Institute for Physical Problems, ul. Kosygina 2, 119334 Moscow, Russia }
\date{\today}

\begin{abstract}
  We present numerical calculations of the snake instability in a
  Fermi superfluid within the Bogoliubov-de Gennes theory of the BEC
  to BCS crossover using the random phase approximation complemented
  by time-dependent simulations. We examine the snaking behaviour
  across the crossover and quantify the timescale and lengthscale of
  the instability. While the dynamic shows extensive snaking before
  eventually producing vortices and sound on the BEC side of the
  crossover, the snaking dynamics is preempted by decay into sound due
  to pair breaking in the deep BCS regime. At the unitarity limit,
  hydrodynamic arguments allow us to link the rate of snaking to the
  experimentally observable ratio of inertial to physical mass of the
  soliton. In this limit we witness an unresolved discrepancy between
  our numerical estimates for the critical wavenumber of suppression
  of the snake instability and recent experimental observations with
  an ultra-cold Fermi gas.
\end{abstract}


\maketitle

\section{Introduction}

Solitons are a ubiquitous feature of fluid dynamics.  In cold gases
they are created in processes of non-equilibrium dynamics
\cite{burgers,phase-imprinting,zutton,engels,becker,shomroni,hamner,
  Yefsah} such as a shock waves, phase and density imprinting,
collisions between condensates, and moving obstacles, or a rapid
quench through a superfluid phase transition \cite{Zurek, Kibble,
  lamporesi}, and may be observed long after the event if they are
sufficiently stable. In strongly correlated Fermi superfluids, solitons
provide a link between hydrodynamics and the poorly understood
dynamics at interatomic length scales.

Dark and gray solitons are solitary wavefronts of reduced density that
are stationary or propagate with a subsonic velocity on a
background. In the context of superfluids, solitons are also called
domain walls, as they are associated with a kink in the superfluid
phase and thus separate domains of different phase. In
weakly-interacting Bose-Einstein condensates (BECs) the study of dark
and gray solitons has begun more than a decade ago
\cite{reinhardt_clark,phase-imprinting,burgers}.  While solitons live
long enough to be observed, they are subject to a dynamical
instability that leads to bending (snaking) of the depletion plane and
eventually to the formation of vortex filaments or vortex rings
\cite{anderson_PRL, zutton}.  This process limits the lifetime
of the soliton as the structure of the initial topological excitation
is lost. The timescale of the decay is given by the excitation
spectrum of this ``snake instability''.

Ultra-cold atomic gases offer the opportunity to study the properties
of solitons during the crossover from a weakly interacting BEC of
pre-formed pairs of fermions to a strongly-correlated superfluid with
unitarity-limited interactions and eventually to a
Bardeen-Cooper-Schrieffer (BCS)-type superfluid with long-range
pairing correlations \cite{giorgini,Ketterle2008}. The crossover is
controlled by the dimensionless parameter $1/(k_F a)$, where $a$ is
the $s$-wave scattering length between Fermi atoms of opposite
(pseudo) spin, $k_F= (3\pi^2n)^{1/3}$ the Fermi wavenumber, and $n$
the density. While soliton properties in the BEC limit $1/(k_F a)\gg
1$ are expected to match those of weakly interacting BECs that are
well described by the Gross-Pitaevskii equation \cite{Brand08,
  Carr08}, the situation is less clear but very interesting in the
crossover region around the unitarity limit where $1/(k_F a) = 0$. The
BCS limit $1/(k_F a) \ll -1$ is currently not accessible to ultra-cold
gas experiments. So far only the very recent experiment of Yefsah {\it
  et al.} has observed dark solitons in the crossover regime
\cite{Yefsah}.  Here, solitons were created by phase imprinting and
their subsequent dynamics in a prolate trap was observed, in order to
obtain data for the ratio of inertial to physical mass.  The solitons
were seen to be remarkably stable against snaking; eventually, the
signatures of the snaking instability appeared for certain trap aspect
ratios.

A standard theoretical approach to modeling the BEC--BCS crossover is
the Bogoliubov-de Gennes (BdG) crossover theory and its extensions
based on diagrammatic many-body theory
\cite{giorgini,spuntarelli}. Since there is no convenient small
expansion parameter at unitarity, where the scattering length diverges
and the particle separation $n^{-1/3}\sim k_F^{-1}$ is the only
available length scale, BdG crossover theory is non-perturbative and
approximate in nature.  The first numerical study of stationary
solitons in 3D fermionic gases in this framework was done by Antezza
\emph{et al.}  \cite{antezza} (see also the work in 1D in
Ref. \cite{1Dbdg}). Subsequent studies found numerical solutions for
moving solitons and investigated their properties
\cite{scott_PRL,liao_brand,scott_brand}, but have been confined to purely
one-dimensional dynamics. Of relevance to this paper are also general
results on solitonic properties outside the approximate BdG crossover
theory based on Landau quasiparticle theory \cite{scott_PRL}, scaling
analysis in the unitarity limit \cite{liao_brand}, and hydrodynamics
\cite{pitaevskii}.

The properties of the system change upon changing the value the
scattering length. In a pure BEC setting, the behavior of the snake
instability is already known, and the excitation spectrum of the decay
has been computed for different confinement potentials
\cite{muryshev,brand_seattle}. More generally, the dispersion relation
of the snaking process has been described in a work by Kamchatnov and
Pitaevskii \cite{pitaevskii} with a hydrodynamical argument. This
method can be also applied to fermionic superfluids, and gives a
prediction for the excitation spectrum of the unstable modes
responsible for the snaking in Fermi gases. Coming from a
hydrodynamical approach, Kamchatnov and Pitaevskii's result is
expected to be valid in the long wavelength limit.

Previous papers \cite{liao_brand, scott_PRL, scott_brand} have
described the stability and the excitation spectrum of travelling dark
solitons in superfluid Fermi gases. Here we wish to perform an
analysis of the decay modes of stationary solitons in the context of
the BdG theory.  We first introduce the analytical argument of
Ref. \cite{pitaevskii} in Sec. \ref{sec:hydro}. Then the excitation
spectrum is studied with comprehensive numerical simulations, upon
using a linear response approach in Sec. \ref{sec:RPA} and a
time-dependent simulations in Sec. \ref{sec:time_evo}.

\section{Hydrodynamic argument} \label{sec:hydro} 
Let us consider a 3D soliton free to move along the $x$
direction. This soliton can be seen as a surface in the Fermi gas with
surface tension $E_s$. The motion happens according to the Newton's
law of motion
\begin{eqnarray} \label{Newton} 
m_s \, \frac{d^2 X}{dt^2} = F_s
\,,
\end{eqnarray} 
where $X(t)$ is the $x$ coordinate of a point on the depletion plane,
and $m_s= 2\,\left.dE_s/d(V^2)\right|_{V=0}$ is the soliton's inertial
mass. At the early stage of the snaking instability, the depletion
plane would bend according to a sinusoidal perturbation
\begin{eqnarray}  \label{cosine} 
X(t) \propto \cos(qy -\omega_q \,t) 
\,.
\end{eqnarray} 
The surface tension appears as a force that tends to minimize the free
energy of the system. More specifically, the force $F_s$ acting on the
soliton at the point $X(t)$ depends on the curvature radius $R$ of the
plane itself
\begin{eqnarray} \label{Fs} 
F_s &=& \frac{E_s}{R} \nonumber\\
R^{-1} &=& \frac{d^2X}{dy^2} 
\,.
\end{eqnarray} 
By substituting Eqs. (\ref{cosine}) and (\ref{Fs}) into
Eq. (\ref{Newton}) one obtains the hydrodynamic approximation for the
instability dispersion \cite{pitaevskii}
\begin{eqnarray} \label{pitaevskii_prediction}
\omega_q = \pm \rmi\,\sqrt{E_s/|m_s|}\,q
\,.
\end{eqnarray}

The value of $\sqrt{E_s/|m_s|}$ can be found analytically at unitarity
by using an argument based on the soliton's oscillation in a harmonic
trap. This oscillation-based approach finds its utility when comparing
the mean-field results to the data coming from experiments.

For this reason, let us call $T_s$ the period of oscillation of the
soliton in this type of potential, and $T_\mathrm{trap}=
2\pi/\omega_\mathrm{trap}$ the inverse of the characteristic frequency
of the trap. The value of $m_s$ can be written in terms of
$T_s/T_\mathrm{trap}$ by using the relation in
Ref. \cite{scott_PRL,liao_brand}
\begin{eqnarray} \label{ms}
|m_s|= m |N_s| \left( \frac{T_s}{T_\mathrm{trap}} \right)^2
\,,
\end{eqnarray} 
where $m$ is the atomic mass. Equation (\ref{ms}) is completely
generic, and it applies to solitons with any interaction strength.

At unitarity it is possible to estimate the energy of the soliton
\cite{liao_brand} $E_s\propto \mu^2$; remembering that $N_s= -\partial
E_s/ \partial \mu$
\begin{eqnarray} \label{Es}
E_s= -\frac{N_s \mu}{2}
\,.
\end{eqnarray} 
By substituting Eqs. (\ref{Es}) and (\ref{ms}) into
Eq. (\ref{pitaevskii_prediction}), after some calculations, we find
\begin{eqnarray} \label{uni_exc}
  |\omega_q| \,\frac{\hbar}{E_F} 
                               &=& \sqrt{\frac{\mu}{E_F}} \left( \frac{T_\mathrm{trap}}{T_s} \right) \frac{q}{k_F}
\end{eqnarray} 
with $E_F= \hbar^2k_F^2 / 2 m $. At unitarity, equation
(\ref{uni_exc}) is an expression that conveniently relates the slope
of the snaking excitation spectrum to the ratio
$T_\mathrm{trap}/T_s$. This ratio (related to the inertial mass ratio
$m_s/m N_s$ by Eq.~(\ref{ms})) is an experimentally measurable
quantity.\\

In the unitarity limit, the BdG theory gives $\mu/E_F\approx
0.6$. In the same limit, both the BdG approach and analytical
arguments \cite{liao_brand} predict a period of oscillation as
$T_s/T_\mathrm{trap}= \sqrt{3}$, therefore
\begin{equation} \label{theo_exc}
  |\omega_q| \,\frac{\hbar}{E_F} 
                  = \sqrt{\frac{\mu}{E_F}}\frac{1}{\sqrt{3}} \frac{q}{k_F}
                               \approx  0.45 \frac{q}{k_F} \,,
\end{equation} 
A different result is found if one assumes that the above hydrodynamic 
argument also applies to the recent experiment of Ref.~\cite{Yefsah}. The observed period of 
oscillation is much larger than the BdG prediction in the whole crossover region, with 
$T_s/T_\mathrm{trap} \approx 14$ at unitarity. Using this value in Eq.~(\ref{uni_exc}), together
with the experimental value $\mu/E_F\approx 0.36$, one obtains
\begin{equation} \label{exp_exc}
  |\omega_q| \,\frac{\hbar}{E_F}  \approx 0.043 \frac{q}{k_F}  \, ,
\end{equation} 
which implies a much slower decay rate.  Even though the rate of
instability, $|\omega_q|$, was not directly measurable in
\cite{Yefsah}, the experiments seem to indicate that solitons having a
long oscillation period have also a very long lifetime against
snaking, in qualitative agreement with Eq.~(\ref{uni_exc}).  The
hydrodynamic arguments also suggests that both effects can be
consistently attributed to a large mass ratio $m_s/(mN_s)$.

\section{RPA theory} \label{sec:RPA}

The excitation spectrum of the snaking instability can be found by
applying a transverse wave perturbation to the system. In the linear
response approach one looks at the behavior of the system at small
times after the perturbation has been applied. Within this approach,
the poles of the static response function correspond to the normal
modes of the system.

The entire excitation spectrum can in principle be found by computing
the static response function $\Pi(\omega_q,q)$ and by looking at the
specific frequencies that make the denominator of $\Pi$ vanish. It is
worth noticing that the unstable modes of the snaking instability
appear at imaginary frequencies.

In the following, we compute the response by using the random phase
approximation (RPA) summation of diagrams \cite{Leggett, bruun,
  minguzzi}, also called ring approximation.

\subsection{Methodology}

A many body system of spin $1/2$ fermions with pairing is described by
the Green's function

\begin{eqnarray}  \label{green1}
G(\mathbf{r},t,\mathbf{r}',t')
\!\! = \!\!
\left(
\begin{array}{cccc}
\langle \hat\psi_\uparrow(\mathbf{r},t) \hat\psi^\dagger_\uparrow(\mathbf{r}',t')  \rangle
                  & \langle \hat\psi_\uparrow(\mathbf{r},t) \hat\psi_\downarrow(\mathbf{r}',t')  \rangle\\
\langle \hat\psi^\dagger_\downarrow(\mathbf{r},t)\, \hat\psi^\dagger_\uparrow(\mathbf{r}',t') \rangle
                  & \langle \hat\psi^\dagger_\downarrow(\mathbf{r},t)\, \hat\psi_\downarrow(\mathbf{r}',t') \rangle \\
\end{array}
\right) 
,~~
\end{eqnarray}
where $\hat\psi_\sigma$ ($\hat\psi_\sigma^\dagger$) is the destruction (creation)
operator for the fermionic species $\sigma= \downarrow,\uparrow$.

In the Bogoliubov-de Gennes (BdG) theory the order parameter
$\Delta(\mathbf{r},t)$ and the density $n(\mathbf{r},t)$ can be
expressed in terms of the Bogoliubov amplitudes $u(\mathbf{r},t)$ and
$v(\mathbf{r},t)$
\begin{eqnarray}
\Delta &=& -g_\mathrm{eff} \sum_j u_j\,v_j^* ,\\
n &=& 2\sum_j v_j\,v_j^*,  
\end{eqnarray}
where the interaction strength $g_\mathrm{eff}$ is given by the
renormalized value \cite{giorgini}
\begin{equation} \label{potential}
\frac{1}{g_\mathrm{eff}} = \frac{m\,k_F}{4\pi\hbar^2} \frac{1}{k_F\,a} - \frac1V \sum_\mathbf{k} \frac{m}{\hbar^2k^2}
\, .
\end{equation}
The value of $a$ is the 3D scattering length describing the
interactions between particles with different spins.

The BdG theory gives an explicit form of the Green's function using
the Bogoliubov amplitudes $u$ and $v$ \cite{PieriStrinati2003}
\begin{eqnarray} 
G(\mathbf{r},\mathbf{r}',\omega_n)
&=&
 \sum_j \frac1{{\rm i}\omega_n-E_j/\hbar} 
  \left( 
    \begin{array}{c}
      u_j(\mathbf{r})\\
      v_j(\mathbf{r})
    \end{array}
  \right)
  \,
  \left( 
      u_j^*(\mathbf{r}'),
      v_j^*(\mathbf{r}')
  \right)
\nonumber\\
  &+&
  \sum_j \frac1{{\rm i}\omega_n+E_j/\hbar} 
\!\!
  \left( 
    \begin{array}{c}
      -v_j^*(\mathbf{r})\\
      u_j^*(\mathbf{r})
    \end{array}
  \right) 
  \!\!\,\!\!
  \left( 
      -v_j(\mathbf{r}'),
      u_j(\mathbf{r}')
  \right)
\,,~~~
\end{eqnarray}
where $\omega_n = (2n+1) \pi/\beta\hbar$ ($n$ integer) is a fermionic
Matsubara frequency. The static problem for the Bogoliubov
amplitudes is solved by finding the solutions of the equations
\begin{equation} 
\left( \begin{array}{cc} 
                    -\frac{\hbar^2\nabla^2}{2m}-\mu & \Delta\\ 
                    \Delta^* & \frac{\hbar^2\nabla^2}{2m}+\mu \\ 
                   \end{array} 
            \right)
\,
\left( \begin{array}{c} 
       u_j\\
       v_j
      \end{array} 
\right)
= E_j \,
\left( \begin{array}{c} 
       u_j\\
       v_j
      \end{array}
\right)\,,
\end{equation}
where $E_j$ are the excitations energies of the Bogoliubov amplitudes.

We are interested in the response of the pair fluctuation
$\hat\psi_\downarrow(\mathbf{r}t) \hat\psi_\uparrow(\mathbf{r}t)$. For
this purpose, let us define
\begin{eqnarray} \label{BCS_vector}
\hat\chi(\mathbf{r},t)  &=&  \hat\psi_\downarrow(\mathbf{r},t) \hat\psi_\uparrow(\mathbf{r},t) \nonumber \\
\hat\chi^\dagger(\mathbf{r},t) &=&  \hat\psi^\dagger_\uparrow(\mathbf{r},t) \hat\psi^\dagger_\downarrow(\mathbf{r},t)
\,,
\end{eqnarray}
and introduce the vector
\begin{eqnarray} \label{vector}
\mathbf{\xi}(\mathbf{r},t) =
\left(
\begin{array}{c}
\langle  \hat\chi(\mathbf{r},t)  \rangle       \\   
\langle  \hat\chi^\dagger(\mathbf{r},t)  \rangle
\end{array}
\right)
\end{eqnarray}
The time-dependence of the vector in Eq. (\ref{BCS_vector}) can be
studied in the linear approximation. In order to study the snaking
instability we apply a small transversal perturbation to the depletion
plane of a soliton, which has the form
\begin{eqnarray} 
 e^{\rmi\,\mathbf{q_y}\,\mathbf{y}+\rmi\,\mathbf{q_z}\,\mathbf{z}}
\,
\mathbf{\phi}_0
\,,
\end{eqnarray}
with $\mathbf{\phi}_0$ a two-component vector. The result in the RPA
approximation is \cite{bruun,fetter}
\begin{eqnarray} \label{response}
\delta\mathbf{\xi}(x,q_y,q_z,t) 
&=&  \int \!\!d\omega e^{\rmi\,\omega\,t}\!\!\int \!\!dx'\,\Pi^\mathrm{RPA}(x,x',q_y,q_z,\omega) \! \mathbf{\phi}_0
\,,~~~~
\end{eqnarray}
where $x$ is the coordinate on the depletion plane of the
soliton, and $q_y,q_z$ are the wavenumber of the transversal
perturbation. The response function is
\begin{eqnarray} \label{piring}
\Pi^\mathrm{RPA}(x,x',q_y,q_z,\omega)
&=& 
\int dx''
\left[ 
  \mathbf{1}_{x,x''} - g_\mathrm{eff}\,\Pi_0(x,x'',q_y,q_z,\omega)
\right]^{-1}
\nonumber\\
&~&~~ 
\Pi_0(x'',x',q_y,q_z,\omega)\,,\nonumber\\
\end{eqnarray}
with $g_\mathrm{eff}$ being the renormalized interaction,
\begin{eqnarray}
\mathbf{1}_{x,x''}
=
\left(
\begin{array}{cccc}
\delta(x-x'') & 0 \\
0  & \delta(x-x'')
\end{array}
\right)
\,,
\end{eqnarray}
and
\begin{eqnarray} \label{matrix1}
\Pi_0
=
\left(
\begin{array}{cccc}
\langle \hat\chi\, \hat\chi^\dagger \rangle & \langle \hat\chi\, \hat\chi \rangle  \\
 \langle \hat\chi^\dagger\, \hat\chi^\dagger \rangle &\langle \hat\chi^\dagger\, \hat\chi \rangle
\end{array}
\right)
\,.
\end{eqnarray}
Here, expectation values are evaluated as averages corresponding to
sums over the quasiparticle amplitudes calculated with the stationary
BdG equations. The same expression can be given in terms of the
Green's function of Eq. (\ref{green1})
\begin{eqnarray}
\Pi_0
=
\left(
\begin{array}{cccc}
-G_{00}\,G_{11}' & -G_{01}\,G_{01}' \\
-G_{10}\,G_{10}' & -G_{11}\,G_{00}'
\end{array}
\right)
\,,
\end{eqnarray}
which can eventually be expressed by using the Bogoliubov
amplitudes. The details of this calculation are presented in Appendix
A.

Finally, the poles of the system are given by the condition
\begin{equation} \label{denom_ring}
\mathrm{det}
\left(
  \mathbf{1}_{x,x''} - g_\mathrm{eff}\,\Pi_0(q,\omega)_{x,x''}
\right)
=0\,,
\end{equation}
with $\Pi_0(q,\omega)$ being a tensor in the coordinates $x$ and
$x''$. The resonant energies of the system are given by the poles of
$\Pi^\mathrm{RPA}$.

These poles can be complex. For definiteness, let us use $\Omega$ for
the real part and $\gamma$ for the imaginary part, so that
\begin{eqnarray}
\omega= \Omega + \mathrm{i} \gamma\,.
\end{eqnarray}
For each value of $q_y$ and $q_z$ these poles must satisfy the relation
\begin{eqnarray} \label{determinant}
\mathrm{det}
\left[ 
\mathbf{1}\,\lambda - g_\mathrm{eff}\,\Pi_0(q_y,q_z,\Omega_{q_y,q_z}+ \mathrm{i}\,\gamma_{q_y,q_z})
\right] = 0
\,,
\end{eqnarray}
with $\lambda=1$. Notice that, for given frequency
$\Omega_{q_y,q_z}- \rmi\,\gamma_{q_y,q_z}$, $g_\mathrm{eff}\,\Pi_0$ is a
matrix in the $2$-dimensional space defined in Eq. (\ref{vector}), so
that Eq. (\ref{determinant}) is equivalent to finding a vector
$|\lambda\rangle$ such that
\begin{eqnarray} \label{eig}
g_\mathrm{eff}\,\Pi_0(q_y,q_z,\Omega_{q_y,q_z}+ \mathrm{i}\,\gamma_{q_y,q_z})| \lambda \rangle = |\lambda\rangle
\,.
\end{eqnarray}
In other words, it is possible to diagonalize $g_\mathrm{eff}\,\Pi_0$
and search for the eigenvalues equal to $1$ for each frequency
$\Omega_{q_y,q_z}+ \rmi\,\gamma_{q_y,q_z}$. The eigenvectors
corresponding to the eigenvalue $1$ are the ones that provide the
resonances.
~\\

\begin{figure}[tbp]
\centering
\includegraphics[width=3in, clip=true]{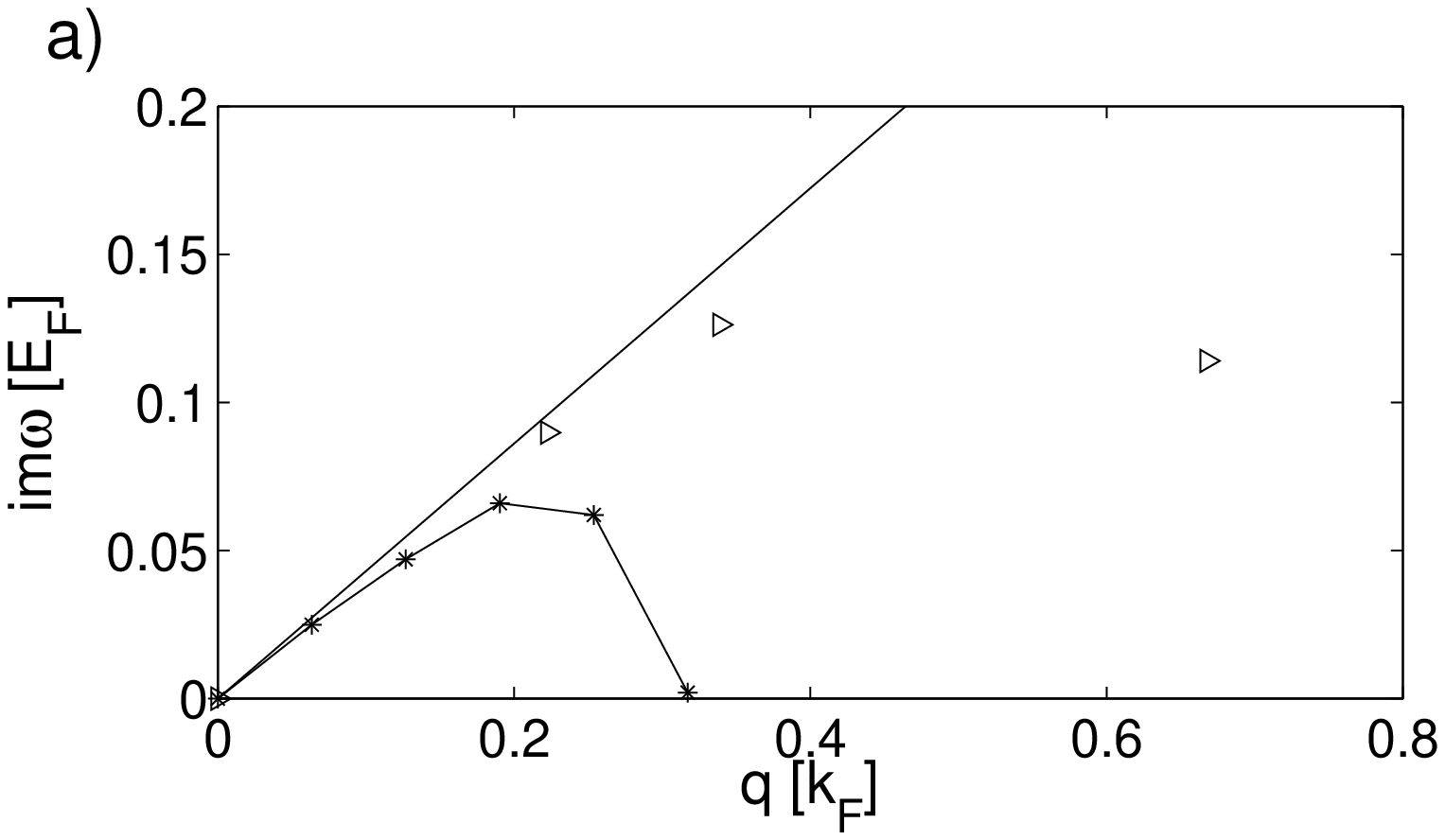}~\\
\includegraphics[width=3in, clip=true]{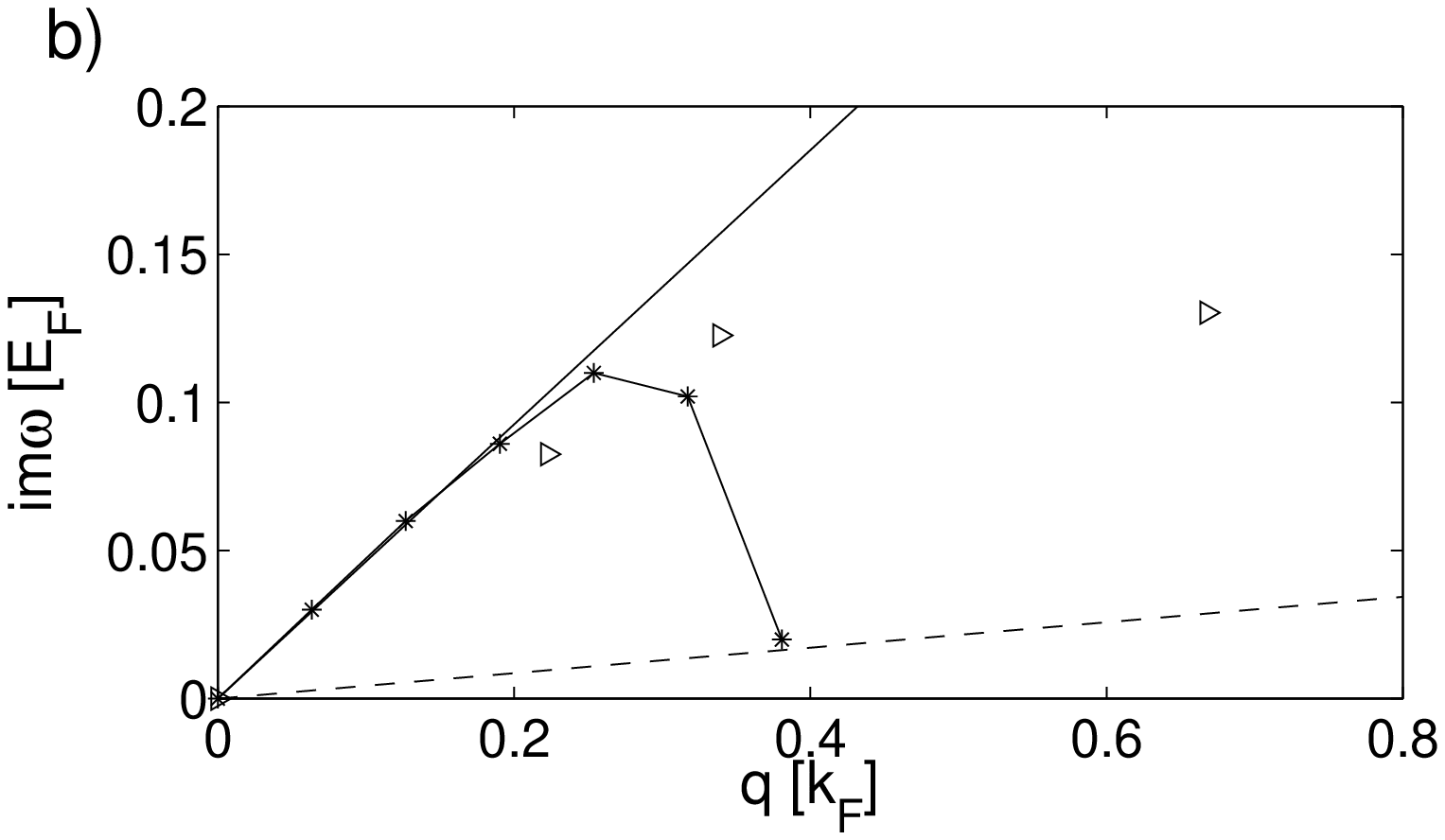}~\\
\includegraphics[width=3in, clip=true]{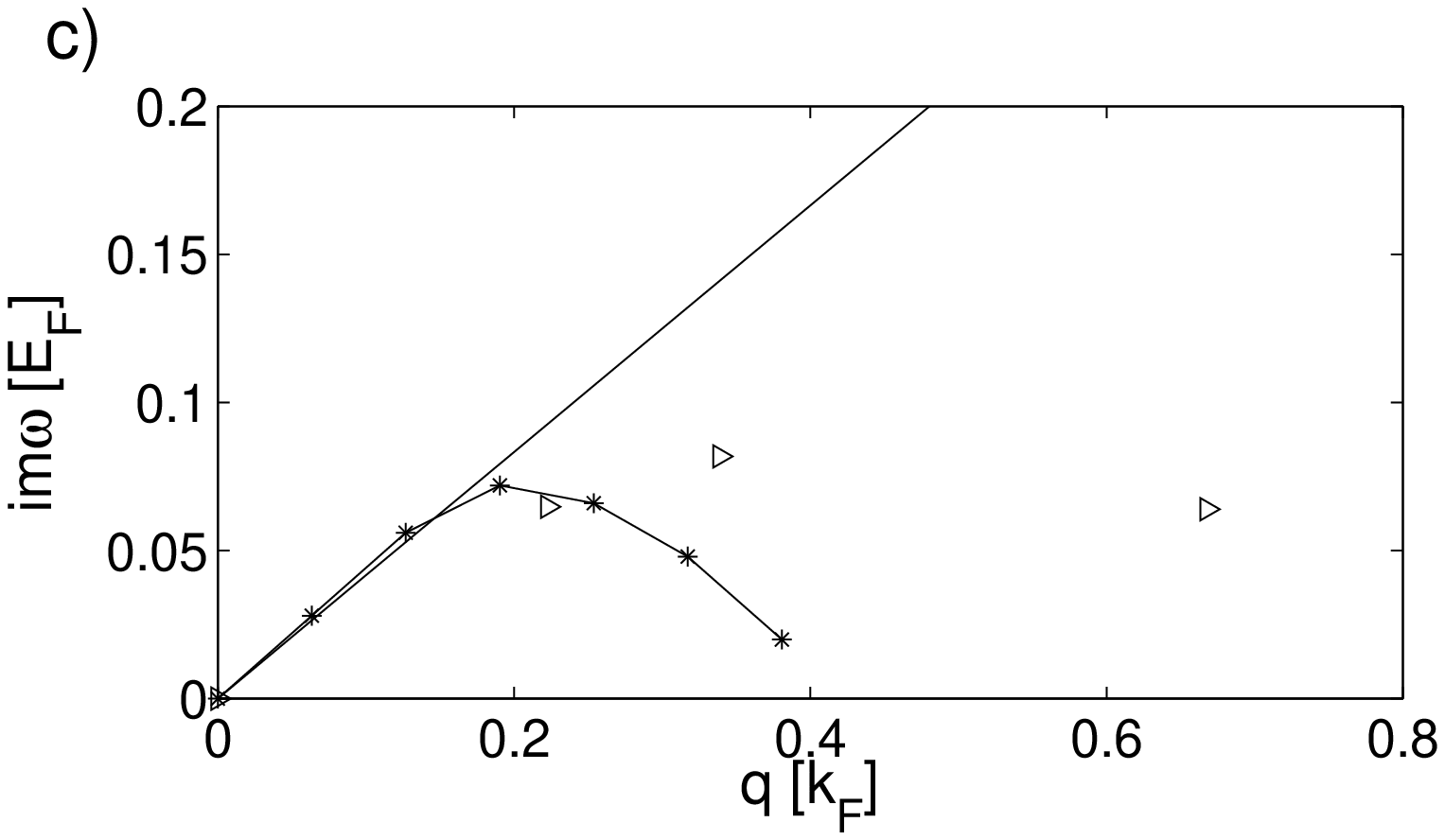}~\\
\includegraphics[width=3in, clip=true]{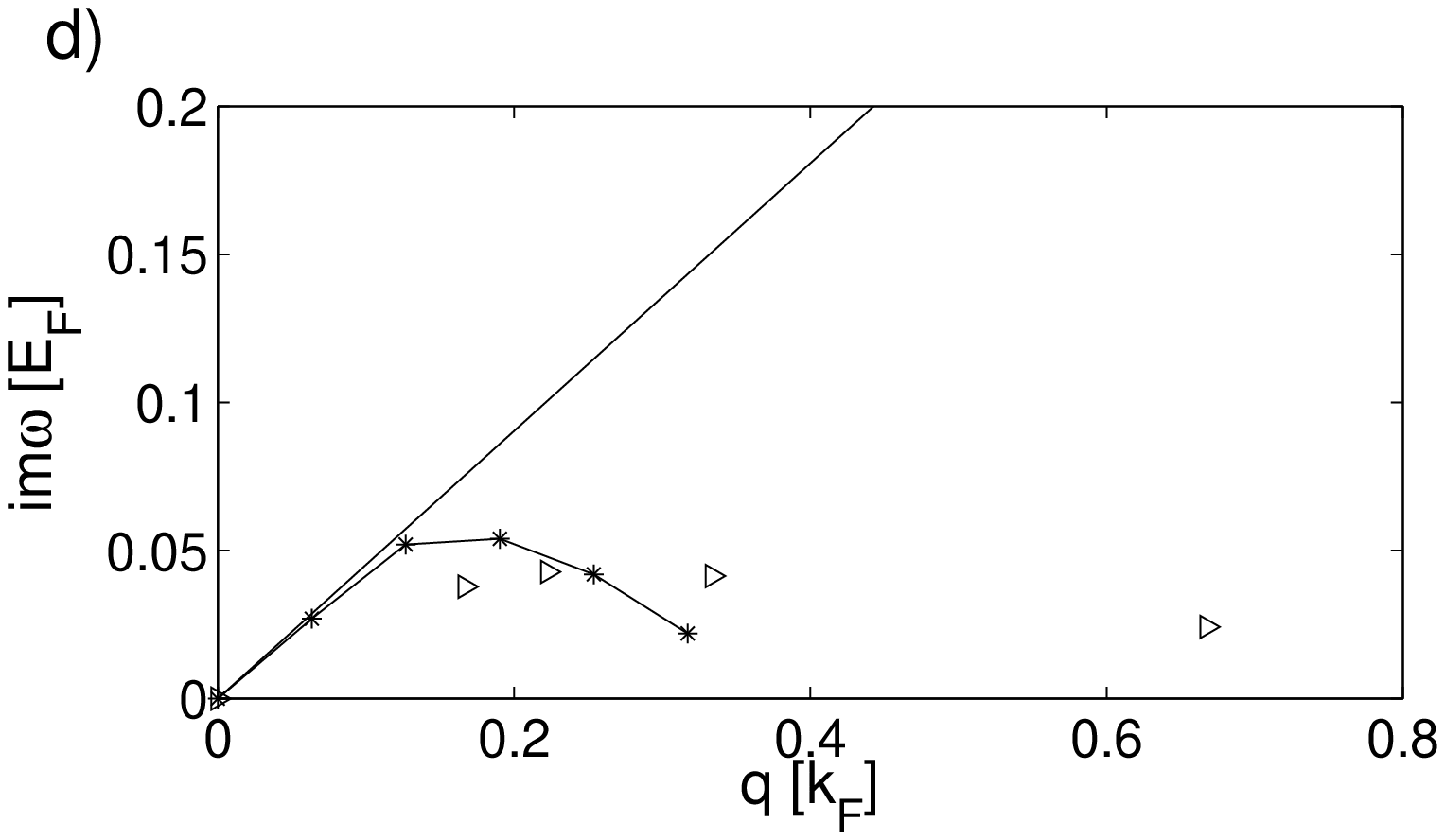}~\\
\caption{ Rate of snake instability of a dark soliton as 
a function of the wavevector of the snaking oscillation. Stars joined by 
solid lines correspond to the resonant poles calculated with RPA 
for $1/(k_F\,a)=0.2 (a), 0 (b), -0.5 (c),  -0.75 (d)$.   The numerical error 
for these points is about $0.001\,E_F$. Triangles correspond to the growth rate 
obtained in  the time-dependent BdG simulations.  The solid straight line in each 
panel is the hydrodynamic prediction (\protect\ref{pitaevskii_prediction}), valid 
in the small $q$ limit, with $E_s$ and $m_s$ obtained from the stationary BdG 
equations. At unitarity (panel (b)) this line coincides with Eq.~(\protect\ref{theo_exc}), 
while the dashed line represents Eq.~(\protect\ref{exp_exc}), which is the same 
hydrodynamic relation, but using the experimental value for the chemical 
potential and the period of oscillation of solitons as measured in \cite{Yefsah}.} 
\label{fig:RPA_poles}
\end{figure}

\subsection{Results}

For plotting the snake instability spectrum we seek to find the
complex values for small values of $q$. In the RPA analysis we see
that the spectrum is purely imaginary ($\Omega=0$) close to $q=0$. The
results are plotted in Fig. \ref{fig:RPA_poles} for various values of
the interaction $1/(k_F\,a)$. 

The search algorithm discretizes the plane and looks at different
combinations of energy and $q$; when a unit eigenvalue ($\lambda=1$)
is found we plot the corresponding resonance value. At our highest
resolution, we have chosen an energy step of $0.001\,E_F$ and a step
for $q$ of $0.065\,k_F$. We have chosen the resolution in the energy
as the error in the determination of the poles. In these calculations
we have set an energy cutoff of $E_C= 20\,E_F$; larger cutoffs would
imply extremely time consuming calculations.

We point out that a cutoff effect in the RPA calculations seems to be
relevant on the BEC side of the crossover. As a consequence, for
$1/(k_F\,a)> 0.2$ our RPA results do not show any excitation poles in
the range predicted by the hydrodynamical argument, and convergence
with the results in Ref. \cite{muryshev} is not obtained.

~\\
~\\
\begin{center}  
\begin{tabular}{c|c|c}
 &  \multicolumn{2}{c}{\centering$\left|\mathrm{Im}(\omega_q)\right|/q  ~~ [v_f]$}\\
\hline
$1/k_F a$ & Hydrod. & RPA\\
\hline
0.2	&0.215 & $0.2  \pm 0.01$\\
0	&0.232 & $0.23 \pm 0.01$\\
-0.5	&0.208 & $0.22 \pm 0.01$\\
-0.75	&0.226 & $0.21 \pm 0.01$
\end{tabular}
\end{center}
~\\
{ \small
Table 1: Slope of the dispersion law of the unstable excitation mode for 
different interaction strengths. The hydrodynamic prediction corresponds 
to the ratio $E_s/m_s$ as in Eq.(4); this ratio is computed by using the 
method of Ref.\cite{liao_brand}. The RPA values are calculated from the 
numerical RPA results in Fig. \ref{fig:RPA_poles}, in the low $q$ limit.
~\\~\\
}  
~\\
~\\

At low $q$ the RPA results are in agreement with
Eq. (\ref{pitaevskii_prediction}), as shown in Table 1. This is a
nontrivial result, and we stress this is the first microscopic
numerical check of the hydrodynamic argument by Kamchatnov and
Pitaevskii. The RPA results deviate from the linear slope downward at
large $q$.

\section{Time-dependent simulations} \label{sec:time_evo} 

The time-dependent Bogoliubov-de Gennes (TDBdG) equations are
numerically solved to further study the snaking instability. We
simulate the time evolution of the soliton for a set of values of the
interaction strength in the crossover. As discussed previously, the
soliton itself is an unstable solution of the stationary BdG
equations; for this reason the snaking has to be induced by applying a
small initial perturbation to the system.

\subsection{Methodology}

The functions $u$ and $v$ solve the equation of motion \cite{Challis}
\begin{equation} \label{BdG}
\left( \begin{array}{cc} 
                    -\frac{\hbar^2\nabla^2}{2m}-\mu & \Delta\\ 
                    \Delta^* & \frac{\hbar^2\nabla^2}{2m}+\mu \\ 
                   \end{array} 
            \right)
\,
\left( \begin{array}{c} 
       u_j(\mathbf{r},t)\\
       v_j(\mathbf{r},t)
      \end{array} 
\right)
= \rmi\,\hbar\partial_t
\left( \begin{array}{c} 
       u_j(\mathbf{r},t)\\
       v_j(\mathbf{r},t)
      \end{array}
\right)\,.
\end{equation}
We choose to confine the system in a box with periodic boundary
conditions along the transverse directions $y$ and $z$ and Dirichlet
boundary conditions in the longitudinal $x$ direction. We keep the box
size in the $x$ and $z$ directions fixed ($30 k_F^{-1}$ and $10
k_F^{-1}$, respectively) while varying the size in the transverse $y$
direction, $L_y$, in the range from $10$ to $40 k_F^{-1}$; this is the
direction along which we perturb the soliton to obtain snaking. The
regularization procedure needed to remove the ultraviolet divergences
in the BdG equations is the same as the one used before in the RPA
method. In the time-dependent simulations we use the cutoff energy
$E_c= 50 E_F$ for positive values of the scattering length and $E_c=
30 E_F$ for negative values.

We first prepare the soliton as a stationary solution of the BdG
equations and then we modify it by imposing a tiny phase shift
$\delta\phi = 0.02 \pi \sin(2\pi \, y/L_y)$ on $\Delta$ at the left of
the soliton plane (for $x<0$). This slightly perturbed state is used
as the initial state of the TDBdG simulation, at $t=0$.  The small
perturbation acts as a seed for the snaking of the soliton. The
characteristic wavenumber of the perturbation is $q = 2\pi/L_y$. The
position $x(t)$ of the nodal plane is then measured at $y=L_y/2$ and
$z=0$, and fitted with the exponential law $x(t)\propto
\exp(\gamma\,t)$.

\begin{figure}[tbp]
\centering
\includegraphics[width=3.4in,clip]{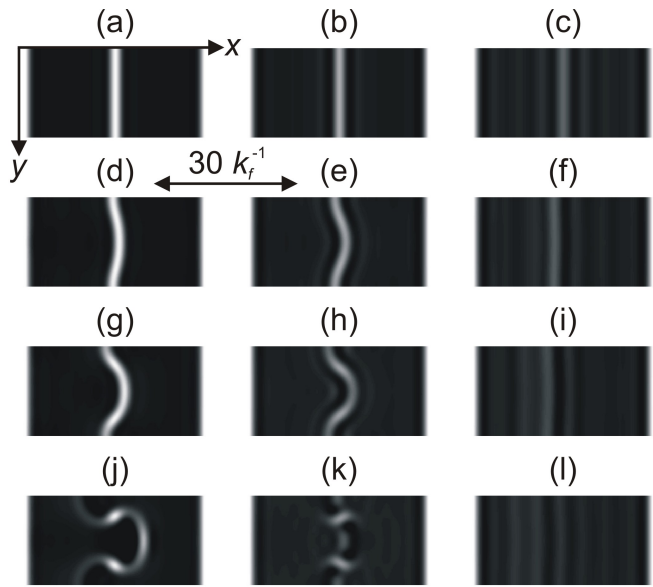}
\caption{Decay of the soliton for various values of the interaction
  strength $1/k_Fa$ in the BCS regime. The column on the left depicts
  the evolution of the soliton at unitarity, the middle column is about 
  $1/k_F\,a=-0.5$ while the rightmost column describes the case $1/k_F\,a=-1$.
  The time evolution at unitarity is presented at the times $0 \hbar/E_F$ (a), 
  $34 \hbar/E_F$ (d), $43 \hbar/E_F$ (g) and $54 \hbar/E_F$ (j); for the middle 
  column the times are $0 \hbar/E_F$ (b), $67 \hbar/E_F$ (e), $79 \hbar/E_F$ (h) 
  and $90 \hbar/E_F$ (k), while at $1/k_F\,a=-1$ the evolution is shown for
  $0 \hbar/E_F$ (c), $157 \hbar/E_F$ (f), $180 \hbar/E_F$ (i) and $200 \hbar/E_F$ (l).}
\label{fig:soliton_decay}
\end{figure}

\subsection{Results}

The deformation of the soliton grows exponentially at short times and
eventually cause the soliton to decay into vortices, as shown in Fig
\ref{fig:soliton_decay} \cite{bulgac_vortices}. The values of $\gamma$
are given in Fig.  \ref{fig:RPA_poles} as a function of $q$ and for
different values of the interaction strength. In the long wavelength
limit, the TDBdG points approach the hydrodynamic law (4), while
bending downward at larger $q$, similarly to the previous RPA
results. The two theories differ in the way they deviate from the
linear slope. Near unitarity, the RPA calculations seem to better
agree with the hydrodynamical approach than TDBdG. An explanation
might lay in the use of Dirichlet boundary conditions in the
time-dependent approach.  Indeed we have checked that using a box with
hard walls instead of periodic boundary conditions can lower the slope
of the excitation spectrum also in RPA calculations. 

This effect, however, does not explain the large discrepancy between
the RPA and TDBdG at large $q$, where the RPA points bend down much
faster. On the one hand, this could be due to the role of the cutoff
energy $E_c$, which is smaller in the RPA calculations than in TDBdG,
hence limiting the convergence toward cutoff-independent results. 
On the other hand, the discrepancy at large q may be related to
nonlinear effects present in the TDBdG calculation but absent in RPA,
which is a linear response theory.

It is worth noticing that, for $1/k_F\,a=-1$, the TDBdG simulations
give no evidence of snaking instability for any value of $q$, the
result of the evolution being only the emission of phonons, as shown
in the right column of Fig.\ref{fig:soliton_decay}. This can be
explained by considering the decay process discussed in
Ref. \cite{scott_brand}: on the BCS side of the crossover, the soliton
can decay due to pair-breaking when moving at a speed larger than a
critical one, eventually emitting phonons. The critical velocity
becomes vanishingly small in the BCS limit. For our simulation this
implies that the motion induced by the initial phase shift reaches
soon the condition of critical velocity, before developing the snaking
instability.

\section{Discussions and conclusions}

Our calculations assume the system to be uniform in the transverse
direction. In trapped gases, the snaking instability can be suppressed if
the superfluid is tightly confined in the transverse direction. It makes
sense to calculate the effect of the transverse confinement in terms of
the relevant parameter
\begin{equation}
\eta= \frac{\mu}{\hbar\omega_\perp}
\end{equation}
where $\mu$ is the chemical potential and $\omega_\perp$ is the harmonic
trapping frequency.  We consider both the BEC and the unitary Fermi gas in
the Thomas-Fermi approximation, where the density profile is determined by
\begin{equation}
\mu= \frac{1}{2} m \omega^2\rho^2 + \mu_{\rm loc}(n)
\end{equation}
where $\mu_{\rm loc}(n)$ the chemical potential of a uniform gas of
density $n$, fixed by the equation of state, and $\rho$ is the transverse
(radial) coordinate. The Thomas-Fermi radius $\rho_\mathrm{TF}$ is
determined by $ \mu_{\rm loc} = 0$, yielding
$\rho^2_\mathrm{TF}=2\mu/(m\omega^2)$.

In the BEC case, defining the healing length as
$\xi=\hbar/\sqrt{2m\mu_{\rm loc}}$ we can calculate the dimensionless
number of healing lengths across the condensate,
\begin{equation}
N_\xi = 2\, \int_0^{\rho_{\rm TF}} \frac{d\rho}{\xi} = \pi\,\eta
\end{equation}
where the second equality follows by virtue of the Thomas-Fermi
approximation and the details of the equation of state are irrelevant. In
Murychev et al. \cite{muryshev}, the critical value $\eta_c\approx 2.4$
was determined numerically for suppression of the snaking instability for
$\eta < \eta_c$. This corresponds to a value of $N_\xi^c\approx 7.5$.
Numerical calculations in a two-dimensional channel with hard walls give a
similar value,  $N_\xi^c\approx 6$ \cite{brand_seattle}.  These values can
be compared with calculations of the homogeneous and infinite soliton
plane, which has a long-wavelength instability at wavenumber
$k_c=1/\sqrt{2}\,\xi$ \cite{muryshev}, corresponding to a wavelength of
$2\pi/k_c=2\sqrt{2}\approx{8.9}$, close enough to the above values. We
conclude that snaking occurs when a full unstable wavelength fits onto the
transverse Thomas-Fermi profile. 

As shown in Appendix B and Fig. \ref{ch_length}, the healing length is
also a relevant length scale in the crossover up to the unitarity
regime, and seems to set the relevant length scale for
short-wavelength suppression of the snaking instability, although the
available numerical data is not entirely conclusive.

For a unitary Fermi gas the relevant length unit is the inverse Fermi
wave number $k_F^{-1}=(2\pi^2n)^{-1/3}$ and $\xi=0.7\,k_F^{-1}$. Using
the equation of state $\mu_{\rm loc}(n)=
(1+\beta)\,\hbar^2\,k_F^2/2\,m$, we obtain
\begin{eqnarray}
N_{k_F^{-1}}&=& 2\, \int_0^{\rho_{\rm TF}} k_F\,d\rho \nonumber\\
     &=& -\frac{\pi}{\sqrt{1+\beta}}\,\eta \approx 5\,\eta
\end{eqnarray}
where $1+\beta\approx 0.4$ was used \cite{blume}.  The numerical data
we have obtained suggests a critical wavenumber of $0.5$ to $1.0\
k_F$, which would suggest a length scale of $6$ to $12\ k_F^{-1}$, or
a value of $\eta_c^{BdG}\approx 1.5$ to $3$. However, the recent MIT
experiment reports $\eta_c \approx 25$ corresponding to
$N_{k_F^{-1}}\approx 125$.  This is obviously a significant
discrepancy. The solitons are stable in a much wider regime in
experiment than would be expected from the BdG calculations and this
demands further theoretical investigations.

To conclude, we have performed a comprehensive analysis of the snake
instability across the BEC-BCS crossover within the mean-field BdG
approximations, both by using a time-dependent approach and a response
function-based method. In our analysis, we have seen the snake instability
to occur in the crossover, but may be preceded by decay into sound in the
deep BCS regime.

In the long wavelength limit mean-field hydrodynamic arguments predict
that the timescale of the decay is set by the soliton energy and
mass. Our BdG calculations well agree with this prediction. However, for
smaller wavelengths in the BCS regime there is a departure from
this behavior; the departure might be due to pair-breaking or boundary
conditions.

On the other hand, the timescales measured in experiment are much longer.
If these experimental results are confirmed other effects must also be
taken into account to accurately describe the snaking in the crossover
region.

\begin{acknowledgments}
  We thank Sandro Stringari and Martin Zwierlein for insightful
  discussions. AC is grateful to the Wenner-Gren foundations for
  financial support. The work is also supported by ERC through the
  QGBE grant and by Provincia Autonoma di Trento. RS is grateful for
  the use of the AURORA supercomputing facilities in Trento.
\end{acknowledgments}

\section*{Appendix A: The renormalization} \label{sec:appendix_A}

The un-renormalized response function $\Pi^\mathrm{RPA}$ in the RPA
approximation is
\begin{eqnarray}
\Pi^\mathrm{RPA}&~&(x,x',q_y,q_z,\omega) =
\int dx''
\left[ 
  \mathbf{1}_{x,x''} \right. \nonumber\\
  &~&\left. - g\,\Pi_0(x,x'',q_y,q_z,\omega)\right]^{-1}
\Pi_0(x'',x',q_y,q_z,\omega) \nonumber \\ 
\end{eqnarray}
where $g=4\hbar^2\pi\,a_s/m$ the ``bare'' interaction and, $\Pi_0$ is
the lowest order response function given by Eq. (\ref{matrix1}). The
expression for $\Pi_0$ can be evaluated by using the Wick's
theorem. For example, at $T=0$
\begin{widetext}
\begin{eqnarray}
\langle \hat\chi\, \hat\chi^\dagger  \rangle (x,x',q_y,q_z)
&=&
\frac{1}{\beta\hbar}
\sum_{\omega_l,k_x,k_y}
\langle \hat\psi_\downarrow(x,k_x+q_y,k_y+q_z,\omega_n+\omega_l)\,
        \hat\psi^\dagger_\uparrow(x,k_x+q_y,k_y+q_z,\omega_n+\omega_l)\, 
\nonumber\\
&~&~~~~~\times
        \hat\psi^\dagger_\uparrow(x',k_x,k_y,\omega_n)\,
        \hat\psi^\dagger_\downarrow(x',k_y,k_z,\omega_n) \rangle
\nonumber\\
&=&
-G_{00}\,G'_{11}
\nonumber\\
&=&
-\frac{1}{\beta\hbar}
\sum_{\omega_l,k_x,k_y}
\langle \hat\psi_\uparrow(x,k_x+q_y,k_y+q_z,\omega_n+\omega_l)\,
        \hat\psi^\dagger_\uparrow(x',k_x+q_y,k_y+q_z,\omega_n+\omega_l)\rangle
\nonumber\\
&~&~~~~~\times
\langle \hat\psi^\dagger_\downarrow(x',k_x,k_y,\omega_n)\,\hat\psi_\downarrow(x,k_y,k_z,\omega_n) \rangle
\nonumber\\
&=&
-\frac{1}{\beta\hbar}
\sum_{\omega_l,k_x,k_y}
G_{00}(x,x',k_x+q_y,k_y+q_z,\omega_n+\omega_l)\,G_{11}(x',x,k_y,k_z,\omega_n)
\nonumber\\
&=&
-\sum_{\eta,\eta'} 
\Big\{
u_\eta(x)\,u_\eta^*(x')\,u_{\eta'}^*(x')\,u_{\eta'}(x)
\frac{1}{\hbar^{-1}\left(E_{\eta}+E_{\eta'}\right)-\omega + \rmi\,\epsilon}
\nonumber\\
&+&
v_{\eta}(x)\,v_{\eta}^*(x')\,v_{\eta'}^*(x')\,v_{\eta'}(x)
\frac{1}{\hbar^{-1}\left(E_{\eta}+E_{\eta'}\right)+\omega + \rmi\,\epsilon}
\Big\}
\,,
\end{eqnarray}
\end{widetext}
where the index $\eta$ is a vector that contains the information over
the wave vectors and $\omega$:
\begin{eqnarray} 
\eta &=& (k_x,k_y,\omega_n)\nonumber \\
\eta' &=& (k_x+q_x,k_y+q_y,\omega_n+\omega_l)\,.\nonumber \\
\end{eqnarray} 

Equation (\ref{matrix1}) can be rewritten as
\begin{eqnarray} 
\Pi_0
=
\left(
\begin{array}{cccc}
-G_{00}\,G_{11}' & -G_{01}\,G_{01}' \\
-G_{10}\,G_{10}' & -G_{11}\,G_{00}'
\end{array}
\right)
\,,
\end{eqnarray}

where the prime over $G'$ inverts the coordinates 
\begin{eqnarray}
G_{\alpha\beta}&=& G_{\alpha\beta}(x,x',q_y,q_z)\nonumber\\
G'_{\alpha\beta}&=& G_{\alpha\beta}(x',x,q_y,q_z)\,,
\end{eqnarray}
and 
\begin{widetext}
\begin{eqnarray}
G_{\alpha\beta}(x_1,x_2,q_y,q_z)\,G_{\gamma\delta}(x_3,x_4,q_y,q_z)
&=&
\sum_{\eta,\eta'} 
\Big\{
\chi_{\eta}^{(\alpha)}(x_1)\,(\chi_{\eta}^{(\beta)})^*(x_2)\,(\tilde\chi_{\eta'}^{(\gamma)})^*(x_3)\,\tilde{\chi}_{\eta'}^{(\delta)}(x_4)
\nonumber\\
&~&~~~~\times
\frac{1}{\hbar^{-1}\left(E_{\eta}+E_{\eta'}\right)-\omega + \rmi\,\epsilon}
\nonumber\\
&+&
\tilde{\chi}_{\eta}^{(\alpha)}(x_1)\,({\tilde\chi}_{\eta}^{(\beta)})^*(x_2)\,(\chi_{\eta'}^{(\gamma)})^*(x_3)\,\chi_{\eta'}^{(\delta)}(x_4)
\nonumber\\
&~&~~~~\times
\frac{1}{\hbar^{-1}\left(E_{\eta}+E_{\eta'}\right)+\omega + \rmi\,\epsilon}
\Big\}
\,.
\end{eqnarray}
\end{widetext}
with
\begin{eqnarray}
\chi_{\eta}(x)
= \left(
  \begin{array}{c}
    u_{\eta}(x)\\
    -v_{\eta}(x)
  \end{array}
  \right)\,,
\end{eqnarray}
and
\begin{eqnarray}
\tilde\chi_{\eta}(x)
= \left(
  \begin{array}{c}
    v_{\eta}^*(x)\\
    u_{\eta}^*(x)
  \end{array}
  \right)
\,.
\end{eqnarray}
The energy $E_{\eta}$ is the Bogoliubov energy for the state
$\eta$. The index $\eta$ is given by a set of three numbers: the
number of the longitudinal excitation energy and the transverse
excitation numbers $q_x$ and $q_y$. A standard renormalization
procedure prescribes to use the term $G_{11}\,G'_{00}-V^{-1}\sum m/\hbar^2k^2$
instead (and similarly for the
other term). ~\\

Eventually the equation that gives the poles, in the cutoff independent
form, is
\begin{eqnarray} \label{determ}
&~&\!\!\!
\mathrm{det}
\left(
\begin{array}{cc}
 \!-\frac1g \!-G_{00}\,G'_{11}\! +\! \frac1V\sum \frac{m}{\hbar^2k^2}\! & -G_{01}\,G'_{01} \\
 -G_{10}\,G'_{10} & \!-\frac1g \!-G_{11}\,G'_{00}\! +\! \frac1V\sum \frac{m}{\hbar^2k^2}\!
\end{array}
\right) \nonumber\\
&~& ~~~~~~ = 0
\,.
\end{eqnarray}
Notice that Eq. (\ref{determ}) can be rewritten as 
\begin{equation}
\mathrm{det}
\left(
\begin{array}{cc}
 -G_{00}\,G'_{11} - \frac1g_\mathrm{eff} & -G_{01}\,G'_{01} \\
 -G_{10}\,G'_{10} & -G_{11}\,G'_{00} -\frac1g_\mathrm{eff}
\end{array}
\right)
=0
\,,
\end{equation}
which is the expression used in the text in Eq. (\ref{denom_ring}).

\section*{Appendix B}

\begin{figure}[htbp]
\centering
\includegraphics[width=3.1in,clip]{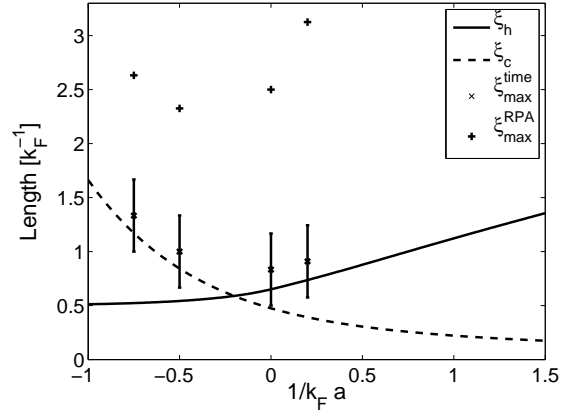}
\caption{ Characteristic lengths in the BCS regime ($\xi_c$) and BEC
  regime ($\xi_h$) as a function of $1/(k_F\,a)$, as computed with Eqs.
  (\ref{xi_coher}) and (\ref{xi_healing}). The values of 
  $\xi^{time}_\mathrm{max}$ and $\xi^{RPA}_\mathrm{max}$ are taken 
  from the time-dependent approach by looking at the longest perturbation wavelength 
  that gives an imaginary spectrum.
}
\label{ch_length}
\end{figure}

As was first shown in Ref. \cite{kuznetsov}, the snake instability is
a long wavelength phenomenon that disappears at shorter wavelengths:
the soliton's imaginary excitation spectrum exists up to a maximum
wavenumber. In a BEC, this critical wavenumber is the inverse of the
healing length. We seek to compare our results to the natural
characteristic lengths of the system in the BEC and BCS regimes.

For a system of boson the natural length is the ``healing length''
\begin{eqnarray} \label{xi_healing}
\xi_h= \frac\hbar{\sqrt{2\,m\,\mu_B}\,,}
\end{eqnarray}
where $\mu_B$ is the boson chemical potential. From Ref. \cite{PieriStrinati2003}
\begin{eqnarray}
\mu_B= \Delta_0 + 2\,\mu_{BCS} \,,
\end{eqnarray}
where $\mu_{BCS}$ is the fermionic chemical potential from the BdG
equations. Therefore
\begin{eqnarray}
\xi_h= \frac{\hbar}{\sqrt{2m \,(\Delta_0 + 2\,\mu_{BCS}) }} \,.
\end{eqnarray}
where $\Delta_0$ is the pairing energy. On the other
hand, the characteristic length in the BCS regime is the ``coherence length''
\begin{eqnarray} \label{xi_coher}
\xi_c= \frac{\hbar\,v_F}{\pi\,\Delta_0} \,,  
\end{eqnarray}

The values for $\mu_{BCS}$ and $\Delta_0$ for the infinite and uniform
as a function of the interaction $1/k_F\,a$ can be found by following
the method in Ref. \cite{Combescot2006}. The result is plotted in
Fig. \ref{ch_length}.

Only in the deep BCS and BEC limit these two quantities become
experimentally relevant. It is however interesting to see that the
characteristic length seems to increase in both limits.


\begin{thebibliography}{10}

\bibitem{burgers} S. Burger, K. Bongs, S. Dettmer, W. Ertmer, and K. Sengstock,
  \emph{Phys. Rev. Lett.}, \textbf{83}, 25 (1999).
\bibitem{phase-imprinting} J. Denschlag, J.E. Simsarian, D.L. Feder,
  Charles W. Clark, L.A. Collins, J. Cubizolles, L. Deng,
  E. W. Hagley, K. Helmerson, W. P. Reinhardt, S.L. Rolston,
  B.I. Schneider, W.D. Phillips Science \textbf{287}, 97 (2000).
\bibitem{zutton} Z. Dutton, M. Budde, C. Slowe, L. Vestergaard Hau,
  \emph{Science} \textbf{293}, 663 (2001).
\bibitem{engels} P. Engels, and C. Atherton, \emph{Phys. Rev. Lett.}
  \textbf{99}, 160405 (2007).
\bibitem{becker} C. Becker et al. \emph{Nat. Phys.} \textbf{4}, 496
  (2008).
\bibitem{shomroni} I. Shomroni, E. Lahoud, S. Levy, J. Steinhauer,
  \emph{Nat. Phys.} \textbf{5}, 193 (2009).
\bibitem{hamner} C. Hamner, J. J. Chang, P. Engels, M. A. Hoefer,
  \emph{Phys. Rev. Lett.}  \textbf{106}, 065302 (2011).
\bibitem{Yefsah} T. Yefsah {\it et al.}, Nature {\bf 499}, 426 (2013).
\bibitem{Zurek} W. Zurek, \emph{Phys. Rev. Lett.}, {\bf 102}, 105702 (2009).
\bibitem{Kibble} T. Kibble, {\it Physics Today}, {\bf 60}, 47 (2007).
\bibitem{lamporesi} G.Lamporesi, S.Donadello, S.Serafini, F.Dalfovo,
  G.Ferrari, \emph{Nat. Phys.} \textbf{9}, 656 (2013).
\bibitem{Bulgac} A. Bulgac, Y.-L. Luo, and K. Roche, \emph{Phys. Rev. Lett.} \textbf{108}, 150401 (2012).



\bibitem{reinhardt_clark} W. P. Reinhardt and C. W. Clark
  \emph{J. Phys. B.} \textbf{30}, L785-L789 (1997).
\bibitem{anderson_PRL} B. P. Anderson, P. C. Haljan, C. A. Regal,
  D. L. Feder, L. A. Collins, C. W. Clark, and E. A. Cornell,
  \emph{Phys. Rev. Lett.} \textbf{86}, 2926 (2001).
  
\bibitem{giorgini} S. Giorgini, L. Pitaevskii, S. Stringari, 
  \emph{Rev. Mod. Phys} \textbf{80}, 4 (2008).
\bibitem{Ketterle2008} W. Ketterle and M. W. Zwierlein,
  \emph{Ultracold Fermi Gases}, Proceedings of the International
  School of Physics "Enrico Fermi", Course CLXIV, Varenna, 20 - 30
  June 2006, edited by M. Inguscio, W. Ketterle, and C. Salomon (IOS
  Press, Amsterdam) 2008, pp. 95-287; \emph{Rivista del Nuovo Cimento}
  \textbf{31}, 247 (2008).

\bibitem{Brand08} J. Brand, L. D.  Carr, and B. P. Anderson, Emergent Nonlinear Phenomena in Bose-Einstein Condensates (2008 Berlin: Springer) chapter 8, p 157.
\bibitem{Carr08} L. D. Carr and J. Brand, Emergent Nonlinear Phenomena in Bose-Einstein Condensates (2008 Berlin: Springer), chapter 7, p 133.
\bibitem{spuntarelli} A. Spuntarelli, P. Pieri, and G.C. Strinati,
  \emph{Physics Reports} \textbf{488}, 111-167 (2010).
\bibitem{antezza} M. Antezza \emph{et al},
  \emph{Phys. Rev. A} \textbf{76}, 043610 (2007).
\bibitem{1Dbdg} J. Dziarmaga, K. Sacha, \emph{arxiv:cond-mat/0407585}.

\bibitem{scott_PRL} R. G. Scott, F. Dalfovo, L. P. Pitaevskii, and S. Stringari
  \emph{Phys. Rev. Lett.} \textbf{106}, 185301 (2011).
\bibitem{liao_brand} R. Liao and J. Brand
  \emph{Phys. Rev. A} \textbf{83}, 041604(R) (2011).
\bibitem{scott_brand} R. G. Scott \emph{et al},
  \emph{New J. Phys} \textbf{14}, 023044 (2012).
\bibitem{pitaevskii} A.M. Kamchatnov and L.P. Pitaevskii
  \emph{Phys. Rev. Lett.}, \textbf{100}, 160402 (2008).


\bibitem{muryshev} A.E. Muryshev \emph{et al.},
  \emph{Phys. Rev. A} \textbf{60}, 4 (1999).
\bibitem{blume} D. Blume, J. von Stecher, and C. H. Greene,
  \emph{Phys. Rev.  Lett.} \textbf{99}, 233201 (2007).
\bibitem{brand_seattle} J.Brand and W.P Reinhardt,
  \emph{Phys. Rev. A} \textbf{65}, 043612 (2002).
\bibitem{Leggett} A. Leggett, \emph{Phys. Rev.} \textbf{147}, 1 (1966)
\bibitem{bruun} G.M. Bruun and B.R. Mottleson,
  \emph{Phys. Rev. Lett.} \textbf{87}, 27 (2001).
\bibitem{minguzzi} A. Minguzzi \emph{et al.},
  \emph{Eur. Phys. J. D} \textbf{17}, 49-55 (2001).
\bibitem{PieriStrinati2003} P. Pieri and G.C. Strinati,
  \emph{Phys. Rev. Lett.}, Vol. \textbf{91}, 030401 (2003).
\bibitem{fetter} A.L. Fetter and J.D. Walecka, \emph{Quantum Field Theory of Many-Particle Systems}, McGraw-Hill (1971)
\bibitem{Challis} K.J. Challis, R.J. Ballagh, and C.W. Gardiner
  \emph{Phys. Rev. Lett.} \textbf{98}, 4 (2007).
\bibitem{bulgac_vortices} The same decay mechanism has been seen in
  the numerical simulations of a Fermi superfluid at unitarity by
  Bulgac et al., \emph{arXiv:1306.4266}. In that case, due to the different
  geometry, the snaking instability produces vortex rings instead of
  straight vortex lines, but the timescale of the process is
  compatible with our results.
\bibitem{kuznetsov} E.A. Kuznetsov and S.K. Turitsyn,
  \emph{Sov. Phys. JETP} \textbf{67}, 1583 (1988).
\bibitem{Combescot2006} R. Combescot, M.Yu. Kagan and S. Stringari,
  \emph{Phys. Rev. A} \textbf{74}, 042717 (2006).


\end{thebibliography}
\end{document}